\documentstyle[12pt]{article}
\begin{document}

\title{CYLINDRICAL COLLAPSE AND GRAVITATIONAL WAVES}
\author{{L. Herrera$^1$\footnote{Postal address: Apartado 80793, Caracas
1080A, Venezuela.}
\footnote{e-mail: laherrera@telcel.net.ve} and N. O.
Santos$^{2,3,4}$\footnote{e-mail: nos@cbpf.br and santos@ccr.jussieu.fr}}\\
\small{$^1$Escuela de F\'{\i}sica, Faculdad de Ciencias,}\\
\small{Universidad Central de Venezuela, Caracas, Venezuela.}\\
\small{$^2$Universit\'e Pierre et Marie Curie - CNRS/FRE 2460,}\\
\small{LERMA/ERGA, Tour 22-12, 4\`eme \'etage, Bo\^{\i}te  142, 4 place
Jussieu,}\\
\small{75252 Paris Cedex 05, France.}\\
\small{$^3$Laborat\'orio Nacional de Computa\c{c}\~ao Cient\'{\i}fica,
25651-070 Petr\'opolis RJ, Brazil.}\\
\small{$^4$Centro Brasileiro de Pesquisas F\'{\i}sicas, 22290-180 Rio de
Janeiro RJ, Brazil.}}
\maketitle

\begin{abstract}
We study the matching conditions for a collapsing
anisotropic cylindrical perfect fluid, and we show that its radial pressure is
non zero on the surface of the cylinder and proportional to the time
dependent part of the field produced by the
collapsing fluid. This result resembles the one that arises for the
radiation - though  non-gravitational - in the spherically symmetric
collapsing dissipative fluid, in the diffusion
approximation.
\end{abstract}

\maketitle

\newpage

\section{Introduction}
Spherical gravitational collapse of a dissipative fluid produces outgoing
radiation which can be modeled with Vaidya spacetime \cite{Santos}. By
using Darmois matching conditions
 \cite{Darmois}, it can be proved that if dissipation within the sphere is
described in the diffusion approximation with heat flux, then the pressure on
the surface of the collapsing
sphere is non zero due to the continuity of the radial flux of momentum
\cite{Santos} (see also
\cite{Bonnor,Herrera} and references therein). Indeed, the fact that heat
flux does not contribute to the fluid radial pressure within the sphere
implies that the pressure does not vanish on the inner part of the boundary surface, because there is
radiation pressure on the outer part of that surface. Of course, in the
streaming out approximation (i.e. when
dissipation within the sphere is described by a radially outgoing null
fluid), the fluid radial pressure is continuous across the boundary
surface, provided the energy density of such
null fluid is continuous too.

It is generally accepted that gravitational waves carry energy, so a
source radiating them should lose mass \cite{Marder,Bondi1,thorne}. On the
other hand such radiation does not
contribute to fluid pressure of the source. Then if one matches a
cylindrical non-dissipative fluid to an exterior containing gravitational
waves, one may expect that
gravitational radiation will exert a non zero pressure on its collapsing
surface, like null radiation induces a non-vanishing radial
pressure on the boundary surface
of a dissipating sphere in the diffusion approximation.

To analyze this problem we studied, by  using the Darmois 
conditions, the matching  of a collapsing
cylindrical non-dissipative anisotropic fluid to a cylindrically symmetric  time dependent exterior  in Einstein-Rosen coordinates
\cite{Einstein}. Matching conditions for LRS spatially homogeneous collapsing dust space time, have been recently considered in \cite{Mena}.

The paper is organized as follows. In the next section we describe the source and its exterior. In Section 3 the Darmois conditions are presented and in Section 4 the results obtained
from its application are exhibited . A specific example is analyzed in Section 5, in order to prove that, in general, the radial pressure does not vanish on the boundary surface of
the source. Finally the results are discussed in the last section.

\section{Collapsing perfect fluid cylinder}

We consider a collapsing cylinder filled with anisotropic non-dissipative
fluid bounded by a cylindrical surface $\Sigma$ and with energy momentum
tensor given by
\begin{equation}
T_{\alpha\beta}^-=(\mu +P_r)V_{\alpha}V_{\beta}+P_rg_{\alpha\beta}+
(P_{\phi}-P_r)K_{\alpha}K_{\beta}+(P_z-P_r)S_{\alpha}S_{\beta}, \label{1}
\end{equation}
where $\mu$ is the energy density, $P_r$, $P_z$ and $P_{\phi}$ are the
principal stresses and $V_{\alpha}$, $K_{\alpha}$ and $S_{\alpha}$ are
vectors satisfying
\begin{equation}
V^{\alpha}V_{\alpha}=-1, \;\; K^{\alpha}K_{\alpha}=S^{\alpha}S_{\alpha}=1,
\;\; V^{\alpha}K_{\alpha}=V^{\alpha}S_{\alpha}=K^{\alpha}S_{\alpha}=0.
\label{2}
\end{equation}
We assume the general time dependent cylindrically symmetric metric
\begin{equation}
ds^2_-=-A^2(dt^2-dr^2)+B^2dz^2+C^2d\phi^2, \label{3}
\end{equation}
where $A$, $B$ and $C$ are functions of $t$ and $r$. To represent
cylindrical symmetry, we impose the following ranges on the coordinates
\begin{equation}
-\infty\leq t\leq\infty, \;\; 0\leq r, \;\; -\infty<z<\infty, \;\;
0\leq\phi\leq 2\pi. \label{4}
\end{equation}
We number the coordinates $x^0=t$, $x^1=r$, $x^2=z$ and $x^3=\phi$ and we
choose the fluid to be comoving in this coordinate system, hence from
(\ref{2}) and (\ref{3})
\begin{equation}
V_{\alpha}=-A\delta_{\alpha}^0, \;\; K_{\alpha}=C\delta_{\alpha}^3, \;\;
S_{\alpha}=B\delta_{\alpha}^2. \label{5}
\end{equation}
The Einstein field equations, $G_{\alpha\beta}=\kappa
T_{\alpha\beta}$ for (\ref{1}), (\ref{3}) and (\ref{5}) reduce to five non
zero components, but we shall need only the following ones
\begin{eqnarray}
G_{11}^-=-\frac{B_{,tt}}{B}-\frac{C_{,tt}}{C}+\frac{A_{,t}}{A}\frac{B_{,t}}{B}+
\frac{A_{,t}}{A}\frac{C_{,t}}{C}-\frac{B_{,t}}{B}\frac{C_{,t}}{C} \nonumber \\
+\frac{A_{,r}}{A}\frac{B_{,r}}{B}+\frac{A_{,r}}{A}\frac{C_{,r}}{C}+
\frac{B_{,r}}{B}\frac{C_{,r}}{C}=\kappa P_rA^2, \label{7} \\
G_{01}^-=-\frac{B_{,tr}}{B}+\frac{B_{,t}}{B}\frac{A_{,r}}{A}+\frac{A_{,t}}{A}
\frac{B_{,r}}{B}-\frac{C_{,tr}}{C}+\frac{C_{,t}}{C}\frac{A_{,r}}{A}+
\frac{A_{,t}}{A}\frac{C_{,r}}{C}=0. \label{10}
\end{eqnarray}

For the exterior vacuum spacetime of the cylindrical surface $\Sigma$ we
take the metric in Einstein-Rosen coordinates \cite{Einstein},
\begin{equation}
ds^2_+=-e^{2(\gamma-\psi)}(dT^2-dR^2)+e^{2\psi}dz^2+e^{-2\psi}R^2d\phi^2,
\label{26}
\end{equation}
where $\gamma$ and $\psi$ are functions of $T$ and $R$ and for the field
equations $R_{\alpha\beta}=0$ we have the gravitational wave field
\begin{equation}
\psi_{,TT}-\psi_{,RR}-\frac{\psi_{,R}}{R}=0, \label{27}
\end{equation}
and
\begin{equation}
\gamma_{,T}=2R\psi_{,T}\psi_{,R},\;\;
\gamma_{,R}=R(\psi_{,T}^2+\psi_{,R}^2). \label{29}
\end{equation}

\section{Junction conditions for the collapsing perfect fluid cylinder}

We take the Darmois junction conditions \cite{Darmois,Bonnor1}, so we suppose
that the first fundamental form which $\Sigma$ inherits from the interior
metric (\ref{3}) must be the same as the one it inherits from the exterior
metric (\ref{26}); and similarly, the inherited second fundamental form
must be the same. The conditions are necessary and sufficient for a smooth
matching without a surface layer.

The equations of $\Sigma$ may be written
\begin{eqnarray}
f_-=r-r_{\Sigma}=0, \label{30} \\
f_+=R-R_{\Sigma}(T)=0, \label{31}
\end{eqnarray}
where $f_-$ refers to the spacetime interior of $\Sigma$ and $f_+$ to the
spacetime exterior, and $r_{\Sigma}$ is a constant because $\Sigma$ is a
comoving surface forming the boundary of the fluid. To apply the junction
conditions we must arrange that $\Sigma$ has the same parametrisation
whether it is considered as embedded in $f_-$ or in $f_+$.

Using (\ref{30}) in (\ref{3}) we have for the metric on $\Sigma$
\begin{equation}
ds^2\stackrel{\Sigma}{=}-d\tau^2+B^2dz^2+C^2d\phi^2, \label{32}
\end{equation}
where we define the time coordinate $\tau$ only on $\Sigma$ by
\begin{equation}
d\tau\stackrel{\Sigma}{=}Adt, \label{33}
\end{equation}
and $\stackrel{\Sigma}{=}$ means that both sides of the equation are
evaluated on $\Sigma$.
We shall take $\xi^0=\tau$, $\xi^2=z$ and $\xi^3=\phi$ as the parameters on
$\Sigma$.

For the exterior metric (\ref{26}) using (\ref{31}) reduces on $\Sigma$ to
\begin{equation}
ds^2\stackrel{\Sigma}{=}-e^{2(\gamma-\psi)}\left[1-\left(\frac{dR}{dT}\right
)^2\right]dT^2
+e^{2\psi}dz^2+e^{-2\psi}R^2d\phi^2, \label{34}
\end{equation}
and is the same as (\ref{32}) if
\begin{eqnarray}
e^{\gamma-\psi}\left[1-\left(\frac{dR}{dT}\right)^2\right]^{1/2}dT\stackrel{
\Sigma}{=}d\tau, \label{35} \\
e^{\psi}\stackrel{\Sigma}{=}B, \label{36} \\
e^{-\psi}R\stackrel{\Sigma}{=}C, \label{37}
\end{eqnarray}
where we assume on $\Sigma$
\begin{equation}
1-\left(\frac{dR}{dT}\right)^2>0, \label{37a}
\end{equation}
so that $T$ is a timelike coordinate.

Equations (\ref{33}) and (\ref{35}-\ref{37}) are the conditions on the
interior and exterior metrics imposed by the continuity of the first
fundamental form on $\Sigma$.

We turn now to the second fundamental form on $\Sigma$. We need the outward
unit normals to $\Sigma$ in $f_-$ and $f_+$; these come from (\ref{30}) and
(\ref{31}) and are
\begin{eqnarray}
n_{\alpha}^-\stackrel{\Sigma}{=}(0,A,0,0), \label{38} \\
n_{\alpha}^+\stackrel{\Sigma}{=}e^{\gamma-\psi}
\left[1-\left(\frac{dR}{dT}\right)^2\right]^{-1/2}\left(-\frac{dR}{dT},1,0,0
\right) \nonumber \\
\stackrel{\Sigma}{=}e^{2(\gamma-\psi)}(-\dot{R},\dot{T},0,0), \label{39}
\end{eqnarray}
where the dot stands for differentiation with respect to $\tau$ introduced
by means of (\ref{35}). Both unit vectors (\ref{38}) and (\ref{39}) are
spacelike provided (\ref{37a}) is satisfied.

The second fundamental form of $\Sigma$ is
\begin{equation}
K_{ab}d\xi^a d\xi^b, \;\; a,b=0,2,3, \label{II11}
\end{equation}
where $K_{ab}$ is the extrinsic curvature given on the two sides by
\begin{equation}
K_{ab}^{\mp}=-n_{\alpha}^{\mp}\left(\frac{\partial^2x^{\alpha}}{\partial\xi^
a\partial\xi^b}+
\Gamma^{\alpha}_{\beta\gamma}\frac{\partial x^{\beta}}{\partial\xi^a}
\frac{\partial x^{\gamma}}{\partial\xi^b}\right). \label{II12}
\end{equation}
The Christoffel symbols are to be calculated from the appropriate exterior
or interior metric, (\ref{3}) or (\ref{26}), $n_{\alpha}^{\mp}$ are given
by (\ref{38}) and (\ref{39}), and $x^{\alpha}$ refers to the equation of
$\Sigma$ in $f_-$ or $f_+$, namely (\ref{30}) or (\ref{31}). The non zero
$K_{ab}^{\mp}$ are as follows
\begin{eqnarray}
K^-_{00}\stackrel{\Sigma}{=}-\frac{A_{,r}}{A^2}, \label{42} \\
K^-_{22}\stackrel{\Sigma}{=}\frac{BB_{,r}}{A}, \label{43} \\
K^-_{33}\stackrel{\Sigma}{=}\frac{CC_{,r}}{A}, \label{44} \\
K^+_{00}\stackrel{\Sigma}{=}e^{2(\gamma-\psi)}\left\{\ddot{T}\dot{R}-\ddot{R
}\dot{T} \right.
\nonumber\\
\left.
-(\dot{T}^2-\dot{R}^2)\left[\dot{R}(\gamma_{,T}-\psi_{,T})+\dot{T}(\gamma_{,
R}-\psi_{,R})\right]\right\}, \label{45} \\
K^+_{22}\stackrel{\Sigma}{=}e^{2\psi}(\dot{R}\psi_{,T}+\dot{T}\psi_{,R}),
\label{46} \\
K^+_{33}\stackrel{\Sigma}{=}-e^{-2\psi}R^2\left(\dot{R}\psi_{,T}+
\dot{T}\psi_{,R}-\frac{\dot{T}}{R}\right). \label{46a}
\end{eqnarray}

The complete junction conditions consist of (\ref{33}) and
(\ref{35}-\ref{37}) together with the continuity of $K_{ab}$ across $\Sigma$.

\section{Results}

In this section, we use the interior and exterior field equations to write
the boundary
conditions in a concise form. In order to do that we first derive some
useful formulae.

From (\ref{35}) we have that
\begin{equation}
e^{2(\gamma-\psi)}(\dot{T}^2-\dot{R}^2)\stackrel{\Sigma}{=}1, \label{50}
\end{equation}
and from (\ref{36}) and (\ref{37})
\begin{equation}
R\stackrel{\Sigma}{=}BC. \label{51}
\end{equation}
Using (\ref{33}) we can differentiate (\ref{51}) yielding
\begin{equation}
\dot{R}\stackrel{\Sigma}{=}\frac{(BC)_{,t}}{A}, \label{52}
\end{equation}
and from the continuity of $K_{22}$ and $K_{33}$ with (\ref{36}) and
(\ref{37}) we obtain
\begin{equation}
\dot{T}\stackrel{\Sigma}{=}\frac{(BC)_{,r}}{A}. \label{53}
\end{equation}
Now differentiating (\ref{52}) and (\ref{53}) with  (\ref{7}), (\ref{10})
and (\ref{33}) we can write the
relation
\begin{eqnarray}
\ddot{T}\dot{R}-\ddot{R}\dot{T}\stackrel{\Sigma}{=}\frac{1}{A^4}\left\{(BC)_{,t}
\left[B_{,t}(AC)_{,r}+C_{,t}(AB)_{,r}\right] \right. \nonumber \\
\left. +(BC)_{,r}\left[+\kappa P_rA^3BC-AB_{,t}C_{,t}-A_{,r}(BC)_{,r}-
AB_{,r}C_{,r}\right]\right\}. \label{55}
\end{eqnarray}
From the continuity of $K_{00}$ and $K_{22}$ and with (\ref{29}),
(\ref{36}), (\ref{50}), (\ref{51}), (\ref{53}) and (\ref{55}) we get
the expression
\begin{equation}
\frac{1}{A^4}\left[(B_{,t}C_{,r}-B_{,r}C_{,t})^2+\kappa P_rA^2(BC)_{,r}^2\right]
\stackrel{\Sigma}{=}(\dot{T}^2-\dot{R}^2)^2\psi^2_{,T}. \label{63}
\end{equation}
Differentiating (\ref{36}) and (\ref{37}) with respect to (\ref{33}) and
(\ref{35}) we obtain
\begin{eqnarray}
e^{2\psi-\gamma}\left[1-\left(\frac{\dot{R}}{\dot{T}}\right)^2\right]^{-1/2}
\psi_{,T}\stackrel{\Sigma}{=}\frac{B_{,t}}{A},
\label{58} \\
e^{-\gamma}\left[1-\left(\frac{\dot{R}}{\dot{T}}\right)^2\right]^{-1/2}
\left(\frac{\dot{R}}{\dot{T}}-R\psi_{,T}\right)\stackrel{\Sigma}{=}\frac{C_{
,t}}{A}. \label{59}
\end{eqnarray}
Then from the continuity of $K_{22}$ and $K_{33}$  with (\ref{58}) and
(\ref{59}) we have
\begin{equation}
\frac{1}{A^2}(B_{,t}C_{,r}-B_{,r}C_{,t})\stackrel{\Sigma}{=}(\dot{T}^2-\dot{
R}^2)\psi_{,T}-\dot{T}\dot{R}\psi_{,R}. \label{60}
\end{equation}
Finally substituting (\ref{60}) into (\ref{63}) and considering (\ref{29})
and (\ref{53})
we obtain
\begin{equation}
\kappa P_{r}\stackrel{\Sigma}{=}e^{2(\psi-\gamma)}\psi_{,R}^2\left(2\frac{\psi_{,T}}
{\psi_{,R}}v-\frac{v^2}{1-v^2}\right), \label{65}
\end{equation}
where $v\stackrel{\Sigma}{=}dR/dT$ denotes the radial velocity of the
boundary surface.
The result (\ref{65})shows that the radial pressure $P_r$
on the surface $\Sigma$ of the collapsing perfect fluid is non zero. The reason for this might be due to the
flux of momentum of the
gravitational wave
emerging from the cylinder. If the cylindrical fluid source is static then
$v\stackrel{\Sigma}{=}0$ and $P_r\stackrel{\Sigma}{=}0$ as expected.

However, it might be argued that $P_r\stackrel{\Sigma}{=}0$ is always
satisfied in cylindrical collapse, leading to an equation for $v$, which
reads
\begin{equation}
v\stackrel{\Sigma}{=}\frac{1}{x}[-1\pm (1+x^2)^{1/2}],
\label{66}
\end{equation}
with $x=4\psi_{,T}/\psi_{,R}$.
By means of an example we shall prove in the next section that this is not
the case.

\section{A pulse solution}
Let us now consider a cylindrical source which is static for a period of time
until it starts contracting and emits a sharp pulse of radiation
travelling outward from the axis. Then, the function
$\psi$ can be written as \cite{carmeli}
\begin{equation}
\psi=\frac{1}{2 \pi}\int_{-\infty}^{T-R}
\frac{f(T^{\prime})dT^{\prime}}{[(T-T^\prime)^2-R^2]^{1/2}} + \psi_{st},
\label{pulse1}
\end{equation}
In (\ref{pulse1}) $\psi_{st}$ represents the Levi-Civita static solution \cite{Levi},
and  $f(T)$ is a function of time representing the strength of the source
of the wave and it is assumed to be of
the form
\begin{equation}
f(T)=f_0\delta (T),
\label{pulse2}
\end{equation}
where $f_0$ is a constant and $\delta (T)$ is the Dirac delta function. It
can be shown that (\ref{pulse1}) satisfies the wave equation (\ref{27}).
Then we get
\begin{eqnarray}
\psi=\psi_{st}, \;\; T<R; \label{pulse3} \\
\psi=\frac{f_0}{2\pi(T^2-R^2)^{1/2}} + \psi_{st}, \;\; T>R. \label{pulse4}
\end{eqnarray}
The function $\psi$,  as well as its derivatives, is regular everywhere except at the wave front determined by the
surface $T=R$, followed by a tail decreasing with $T$.
Using $\psi_{st}=\alpha-\beta\ln R$, with $\alpha$ and $\beta$ being constants,
and (\ref{pulse4})  we obtain
\begin{equation}
x\stackrel{\Sigma}{=}\frac{4 f_0T R}{2\pi \beta(T^2-R^2)^{3/2}-f_0R^2}.
\label{pulse5}
\end{equation}
From (\ref{pulse5}) we see that as $T$ increases, the system tends asymptotically to a static situation, and
for a sufficiently large values of $T$ we have
\begin{equation}
x \stackrel{\Sigma}{\approx} \frac{2f_0 R}{\pi \beta T^2}.
\label{pulse6}
\end{equation}
which is a positive quantity.
On the other hand for values of $T$ larger than, but sufficiently close to $R$, we have that $x$ is negative. This implies that for some value of $T$ (say $T_0$) the denominator in
(\ref{pulse5}) vanishes and $x$ tends to $\mp \infty$ as $T$ goes from $T_0-0$ to $T_0+0$. This in turn implies, because of (\ref{66}) (where only the upper sign before the square root
in (\ref{66}) has to be considered because $v^2<1$), that in the infinitesimal time interval $(T_0-0,T_0+0)$ the velocity $v$ changes from $-1$ to $+1$.
This of course is impossible. Therefore,
condition (\ref{66}) cannot be satisfied in
this example and $P_{r}{\neq}0$ on $\Sigma$.

It is worth noticing that for sufficiently large values of $T$, $\Psi_{,R}
\Psi_{,T}>0$  on $\Sigma$. This quantity, which is related to the rate
of change of the $C$
energy, is shown to be negative in \cite{thorne}. However that proof is
valid only for large values of $R$, which of course is not the case here.

It is also worth noticing that, in general, the static exterior spacetime, i. e.
Levi-Civita spacetime, cannot be matched to a collapsing cylinder with
source (\ref{1}). Indeed, considering that
the exterior spacetime is static, $\psi(R)$, then we have from (\ref{36}),
(\ref{53}) and the continuity of $K_{22}$ that $B\stackrel{\Sigma}{=}B(r)$
and $C\stackrel{\Sigma}{=}C(r)$, which in turn implies because of  (\ref{51})
that $R\stackrel{\Sigma}{=}$ constant showing that the source cannot be
collapsing. This result has been shown before for the particular case of a collapsing dust cylinder \cite{Gutti}.
\section{Conclusion}
The main result of this work is that the radial pressure on the surface of a collapsing anisotropic perfect
fluid cylinder is non zero (\ref{65}). If the system is static we
reobtain the usual result that the radial pressure is zero on the boundary
surface (the same happens in the
slowly evolving case considered by Bondi \cite{Bondi}).

The  physical interpretation of the non-vanishing surface pressure might be
justified through the continuity of the radial flux of momentum across
the boundary surface. However, based on this interpretation what follows is that the collapsing dust cylinder ($P_{r}=0$) should not radiate
gravitational waves, the exterior
spacetime then belongs to the class of time dependent systems without
news, as discussed by Bondi et al in \cite{Bondi1}. Parenthetically Bondi
suggests that pressure-free dust
would be the most clear cut example of such non-radiating time dependent
systems.

Finally,  we can state that if a cylinder with source, as described in (\ref{1}), is collapsing, it will {\it always} produce
gravitational  waves  outside the source, implying the presence of non--vanishing radial pressure on the boundary of the source.

\end{document}